\documentclass[useAMS,usenatbib]{mn2e}

\usepackage{graphicx}

\newcommand{\Msun}{$M_{\odot}$}

\newcommand{\kms}{km\,s$^{-1}$}
\newcommand{\vs}{$v \sin i$}
\newcommand{\teff}{$T_{\rm eff}$}
\newcommand{\lgg}{$\log\,{g}$}

\title[Magnetic field, composition and variability of AR Aur]{Magnetic field, chemical composition and line profile variability of the peculiar eclipsing binary star AR Aur\thanks{
Based on observations obtained at the Canada-France-Hawaii Telescope (CFHT) which is operated by the National Research Council of Canada, 
the Institut National des Sciences de l'Univers of the Centre National de la Recherche Scientifique of France,  and the University of Hawaii.  Also based on observations obtained at the Bernard Lyot Telescope (TBL, Pic du Midi, France) of the Midi-Pyr\'en\'ees Observatory, 
which is operated by the Institut National des Sciences de l'Univers of the Centre National de la Recherche Scientifique of France. }
}

\author[Folsom et al.]{C.P. Folsom$^{1}$\thanks{E-mail: cpf@arm.ac.uk}, O. Kochukhov$^{2}$, G.A. Wade$^{3}$,
J. Silvester$^{3,4}$, S. Bagnulo$^{1}$\\
$^{1}$Armagh Observatory, College Hill, Armagh Northern Ireland BT61 9DG\\
$^{2}$Department of Astronomy and Space Physics, Uppsala University, 751 20 Uppsala, Sweden \\
$^{3}$Department of Physics, Royal Military College of Canada, P.O. Box 17000, Station `Forces', Kingston, Ontario, Canada, K7K 7B4\\
$^{4}$Department of Physics, Engineering Physics \& Astronomy, Queen's University, Kingston, Ontario, Canada, K7L 3N6 }

\begin{document}

\date{Received: 2010; Accepted: 2010}

\pagerange{\pageref{firstpage}--\pageref{lastpage}} \pubyear{2010}

\maketitle

\label{firstpage}

\begin{abstract}
AR Aur is the only eclipsing binary known to contain a HgMn star, making it an ideal case for a detailed study of the HgMn phenomenon.  
HgMn stars are a poorly understood class of chemically peculiar stars, which have traditionally been thought not to possess significant magnetic fields. 
However, the recent discovery of line profile variability in some HgMn stars, 
apparently attributable to surface abundance patches, has brought this belief into question.
In this paper we investigate the chemical abundances, line profile variability, and magnetic field of the primary and secondary of the AR Aur system, 
using a series of high resolution spectropolarimetric observations.  
We find the primary is indeed a HgMn star, and present the most precise abundances yet determined for this star.
We find the secondary is a weak Am star, and is possibly still on the pre-main sequence.  
Line profile variability was observed in  a range of lines in the primary, and is attributed to inhomogeneous surface distributions of some elements.  
No magnetic field was detected in any observation of either stars, with an upper limit on the longitudinal magnetic field in both stars of 100~G. 
Modeling of the phase-resolve longitudinal field measurements leads to a $3\sigma$ upper limit on any dipole surface magnetic field of about 400~G.
\end{abstract}


\section{Introduction}

HgMn stars are chemically peculiar late B stars, characterized by strong overabundances 
of Mn, by up to a thousand times solar, 
and strong overabundances of Hg, by up to a hundred thousand times solar.  
Strong overabundances of Xe, iron-peak elements, Ga, and underabundances of He 
are also often seen in these stars.  
These chemical anomalies are believed to be produced by the selective radiative levitation 
and gravitational settling of chemical elements in the outer stellar layers
\citep*[e.g.][]{Michaud1974}. However,
despite a significant number of observational and theoretical studies, the physical processes responsible for %the origin of 
these strong chemical peculiarities remain poorly understood. 
HgMn stars have been generally thought not to posses strong (dynamically important) magnetic fields
\citep[e.g.][]{Shorlin2002}, setting them apart from magnetic chemically peculiar Ap and Bp stars.  

Recently, line profile variability has been detected in some spectral lines of a number of HgMn stars.  
This has been interpreted as an inhomogeneous surface distribution of 
specific elements (\citealt*{Ryabchikova1999-alphaAnd-abun-var}; \citealt{Adelman2002-HgMn-var-1st,Kochukhov2005-hgmn-inhomo}).  
While magnetic chemically peculiar Ap and Bp stars display a wide range of surface 
abundance inhomogeneities, HgMn stars are the only type of (apparently) non-magnetic A or B stars in which 
such inhomogeneities have been found.  

In Ap and Bp stars, surface abundance inhomogeneities 
are usually attributed to the magnetic field, possibly working together with stellar rotation and magnetically-channelled mass loss.  
This has led some authors to suggest that HgMn stars may have hitherto undetected 
magnetic fields, which could give rise to the observed abundance inhomogeneities 
\citep[e.g.][]{Adelman2002-HgMn-var-1st,Hubrig2006-arAur-lVar}.  
A detailed examination of the spotted HgMn star $\alpha$~And by \citet{Wade2006-alphaAnd} 
found no evidence for such a magnetic field, with an upper limit on the dipole field strength of between 50 and 100 G.  
Notwithstanding the apparent absence of magnetic fields in $\alpha$~And, the more general presence of magnetic 
fields in variable HgMn stars has not been strongly tested.

Doppler imaging of surface abundance inhomogeneities has been performed 
by \citet{Adelman2002-HgMn-var-1st} and \citet{Kochukhov2007-alphaAnd-map-var} for the HgMn star $\alpha$~And.  
Surprisingly \citet{Kochukhov2007-alphaAnd-map-var} find secular evolution of the Doppler maps 
on a time scale of years. This is the only known case of 
surface abundance inhomogeneities evolving with time in any A or B star.  

AR Aurigae (17 Aur, HR\,1728, HD\,34364) is a unique double-lined spectroscopic binary system with nearly identical components (B9V+B9.5V). Being the only eclipsing binary known to host a HgMn star, 
it allows for a precise, model independent determination of fundamental physical parameters.  
The primary of the system has been long known as an HgMn star \citep{Wolff1978}
and line variability in the primary has recently been reported by \citet{Hubrig2006-arAur-lVar}.  
This makes the system an ideal candidate for a more detailed investigation. 

The two main components of the AR Aur system have similar masses of 2.48~\Msun\, and 2.29~\Msun, 
and an orbital period of 4.135 days \citep{Nordstrom1994-arAur-fundamental}.
\citet{Chochol1988-araur-triple} discovered a third component to the AR Aur system, based on the light-time effect.
According to \citet*{Albayrak2003}, the third star has a mass of 0.54~\Msun, a separation of 13~AU, and a period of 23.7 years.
\citet{Nordstrom1994-arAur-fundamental} performed a detailed study of the photometric light curves, deriving radii, 
masses, and effective temperatures of the two brighter, eclipsing components of AR Aur, 
as well as considering the age of the system. They suggest that the secondary is likely a pre-main sequence star
still contracting towards the zero-age main sequence line (ZAMS), while the primary is at the ZAMS.  

\citet{Ryabchikova1998-HgMn-binaries} performed an abundance analysis of both components of AR Aur, 
based on the equivalent widths from photographic spectra reported by \citet{Khokhlova1995-araur-ew}, 
and found strong overabundances of a wide range of elements in the primary, including Mn, Fe, Sr, Y, Pt, 
and Hg, typical of HgMn stars. Inconclusive evidence of the line profile variability in AR Aur A was given
by \citet*{Takeda1979} and \citet*{Zverko1997} based on low-quality photographic spectra.
\citet{Hubrig2006-arAur-lVar} studied AR Aur with modern spectroscopic instrumentation, demonstrating line variability for several elements and suggesting inhomogeneous surface distributions for Sr and Y.  

In this work we search for magnetic fields in AR Aur A and B, making this system the second HgMn 
star with very precise upper limits on its magnetic field. 
Additionally, we derive improved chemical abundances for the primary and the secondary, 
confirming the primary as a prominent HgMn star, and discovering that the secondary is a weak Am star.

\section{Observations}
\label{observations}

Observations were obtained with the Echelle Spectropolarimetric Device for the 
Observation of Stars (ESPaDOnS), a high resolution spectropolarimeter located at the 
Canada France Hawaii Telescope (CFHT), and with NARVAL, a nearly identical instrument 
mounted on the T\'elescope Bernard Lyot at the Observatoire du Pic du Midi, France.  
Both instruments consist of a bench mounted cross-dispersed echelle spectrograph, 
fiber fed from a Cassegrain mounted polarimeter unit.  
They provide near continuous wavelength coverage 
from 3700 to 10500 \AA\, at a resolution of $R=65000$.  
Observations were obtained in spectropolarimetric mode, providing Stokes $V$ spectra 
as well as Stokes $I$ spectra.  The data were reduced using Libre-ESpRIT \citep{Donati1997-major}, 
which performs calibrations and optimal spectrum extraction, 
tailored to the ESPaDOnS and NARVAL instruments.  

A total of 7 spectra, 3 from ESPaDOnS and 4 from NARVAL, were obtained over a period of roughly one month.  
A summary of the observations is presented in Table~\ref{observations_table}.  The orbital phases are calculated
with the ephemeris of \citet{Albayrak2003}, $HJD=2452596.4927+E\times4\fd1346656$, corresponding to the
primary light minimum.

\begin{table}
\centering
\caption{Table of observations. The exposure time includes the number of sub-exposures and the length of a sub-exposure.  
Peak S/N is peak signal-to-noise ratio per CCD pixel. }
\label{observations_table}
\begin{tabular}{ccccc}
\hline \hline \noalign{\smallskip}
HJD & Instrument & Phase & Exposure  & Peak \\
    &            &       & Time (s)  & S/N  \\
\noalign{\smallskip} \hline \noalign{\smallskip}
2454070.87384 & ESPaDOnS & 0.590 & $4\times400$ & 500 \\
2454074.91547 & ESPaDOnS & 0.568 & $4\times200$ & 550 \\
2454076.79173 & ESPaDOnS & 0.021 & $4\times100$ & 400 \\
2454084.53282 & NARVAL   & 0.894 & $4\times400$ & 450 \\
2454089.39584 & NARVAL   & 0.070 & $4\times400$ & 450 \\
2454090.36577 & NARVAL   & 0.304 & $4\times400$ & 350 \\
2454091.39868 & NARVAL   & 0.554 & $4\times400$ & 500 \\
\noalign{\smallskip} \hline \hline
\end{tabular}
\end{table}

\section{Spectral Disentangling}

The spectra of AR Aur exhibit a complex pattern of two very similar absorption line systems, with velocity separation changing
from zero to $\approx$200~\kms\ on the time scale of two days. In addition, one of the components has weak intrinsic line profile
variability. In this situation a spectral disentangling procedure is essential to separate this effect from variable line blending due
to orbital motion of the binary components and to obtain high-quality average spectra for abundance analysis.

We have developed a direct spectral decomposition technique similar to the method described by \citet{Gonzalez2006}. Our algorithm operates as follows. We start with a set of approximate radial velocities for each component and initial guesses of their spectra. Then, the contribution of component B is removed, all spectra are shifted to the rest frame of component A, interpolated on the standard wavelength grid and co-added. This yields a new approximation of the primary spectrum. The same 
procedure is used to update the spectrum of the secondary. This sequence of operations is repeated up to convergence. 
As the second major step, we use a least-squares minimization to derive radial velocities and adjust continuum normalization using low-degree polynomials. Calculation of the spectra and radial velocities is alternated until the changes of these parameters from one iteration to the next are below given thresholds.

\begin{figure*}
\centering
\includegraphics[width=4.0in,angle=90]{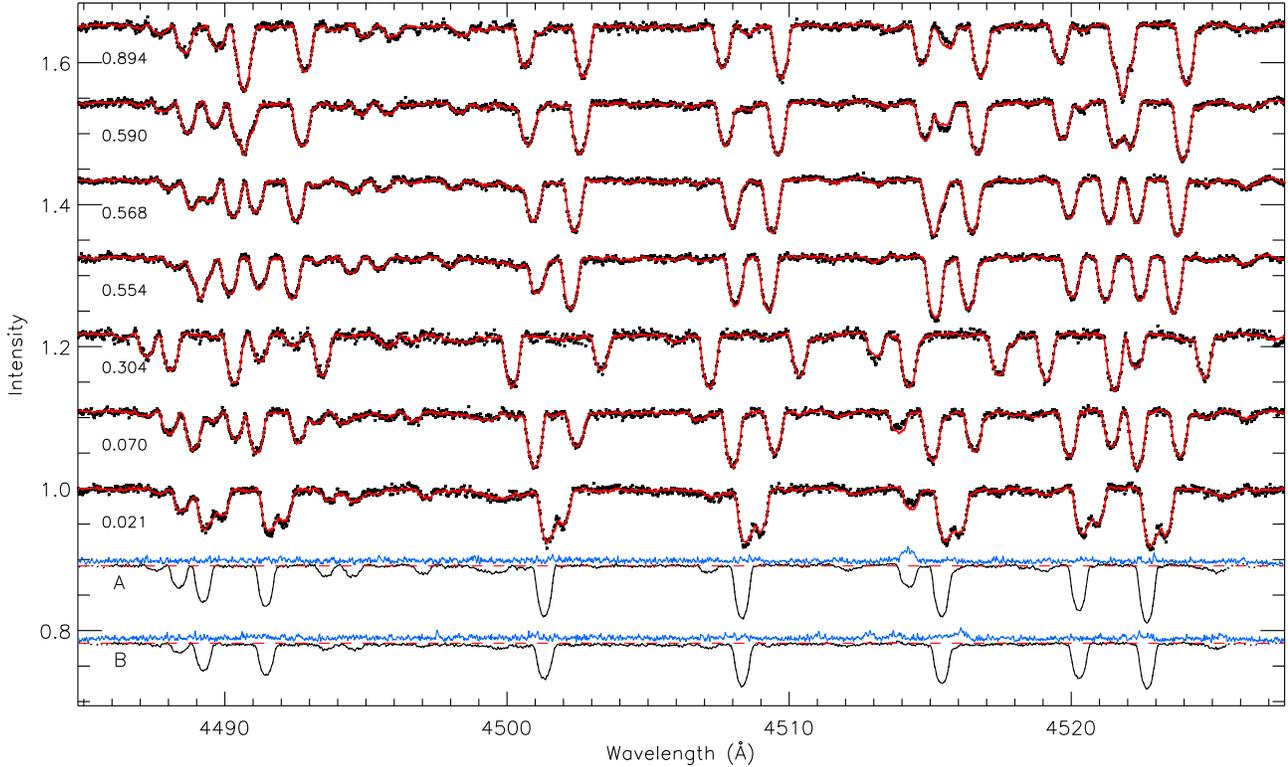}
\caption{Illustration of the spectral disentangling procedure applied to AR Aur. Symbols represent observations
and the thick lines show best fitting spectra. Resulting disentangled spectra of the components A and B are shown 
by the thin lines at the bottom of the figure. The standard deviation spectra (shifted upwards and scaled by a factor of 3) 
diagnose intrinsic line profile variability. An excess of deviation at $\lambda\approx4514$~\AA\ demonstrates
the presence of weak variability for the Pt~{\sc ii} 4514.170~\AA\ line belonging to AR Aur A.}
\label{disent}
\end{figure*}

This spectral disentangling was applied to overlapping 20--70~\AA\ segments of AR~Aur spectra, avoiding Balmer lines and regions affected by the telluric absorption. The final radial velocities calculated by averaging results obtained for different wavelength regions are presented in Table~\ref{rvs}.
Fig.~\ref{disent} illustrates spectral decomposition for the 4485--4525~\AA\ interval, containing the variable Pt~{\sc ii} 4514.170~\AA\ line of AR~Aur~A.

Disentangling yields separate average high-quality ($S/N\approx1000$) spectra of components A and B, 
radial velocities and the standard deviation curves, characterizing the remaining discrepancy between observations and composite model spectra.
The standard deviation is examined in the rest frame of the primary and secondary separately, which allows a straightforward and objective identification of intrinsically variable lines by a coherent excess of standard deviation corresponding to a certain absorption feature (see Fig.~\ref{disent}). 
We found that in all cases the intrinsic spectrum variability is associated with the lines of AR~Aur~A. Thus, to study the spectrum variability of the primary in detail, in the final step of the disentangling procedure we removed from the observations the average spectrum of the secondary and corrected the orbital radial velocity shifts.

\section{Fundamental Parameters}
\label{params}
\begin{figure}
\centering
\includegraphics[angle=90,width=3.3in,height=2.8in]{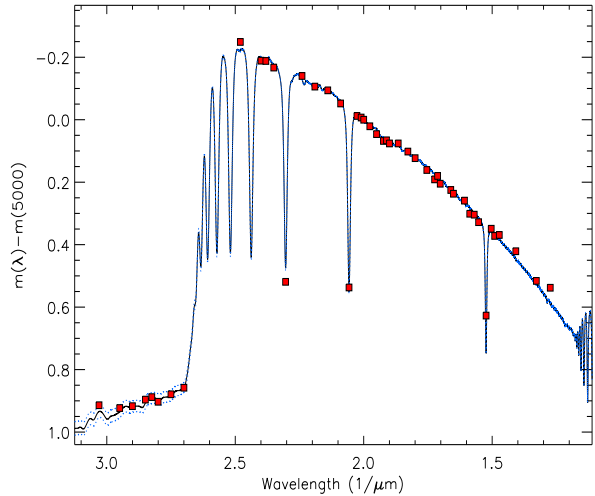}
\caption{Comparison of the observed (symbols) and computed best fit (solid line) composite spectral energy distribution of
AR Aur for the parameters \teff(A)\,=\,10950~K, \teff(B)\,=\,10350~K, and $R_{B}/R_{A}$\,=\,1.033.  
Error bars on the observations are smaller then the symbol size. 
The computed composite spectral energy distributions, with the temperature of the primary varied by $\pm150$~K, 
are shown as a dashed lines. }
\label{flux}
\end{figure}

Effective temperatures for both components of AR~Aur were determined based on the 
spectral energy distribution of the binary system.  
The observed spectrophotometry was taken from the catalogue of \citet{Adelman1989} and synthetic fluxes 
were calculated using the LLModels model atmosphere code \citep{Shulyak2004-llmodels}. 
This code produces plane-parallel, line blanketed model atmospheres in local thermodynamic equilibrium (LTE), 
using an advanced `line-by-line' method for including the effects of metal lines.  
The chemical abundances of \citet{Ryabchikova1998-HgMn-binaries} were used as input for the model atmosphere calculations, 
to account for chemical peculiarities.  
The best fit combination of models gives \teff\,=\,$10950\pm150$~K for the primary 
and \teff\,=\,$10350\pm150$~K for the secondary. 
The $T_{\rm eff}$ uncertainties were conservatively estimated by eye, 
based mainly on the fit to the  Balmer continuum. 
Fig.~\ref{flux} illustrates the agreement between the observed and computed spectrophotometry.
The error bars for the spectrophotometry are smaller then the symbol size in this figure.  

The ratio of radii of the AR Aur system was inferred from
the ratio of Mg~{\sc ii} 4481~\AA\ line equivalent widths, assuming identical Mg abundances.  
This assumption was verified during the abundance analysis presented in Sect.~\ref{Chemical Abundances}.  
The Mg abundances for the two components, based on a wide range of Mg lines, 
are within uncertainty of each other, and nearly solar in both stars.  
We find $R_{B}/R_{A} = 1.033\pm0.005$, which is consistent with the value of $1.020\pm0.015$ 
determined by \citet{Nordstrom1994-arAur-fundamental}.
Our model predicts the central surface brightness 
ratio $J_B/J_A=0.874$ in excellent agreement with the observed mean value 
of $0.870\pm0.006$ in the $V$ and $y$ bands \citep{Nordstrom1994-arAur-fundamental}.

Our temperatures of AR Aur A and B are in agreement with the values of \citet{Nordstrom1994-arAur-fundamental} but, probably, are somewhat more precise.
While those authors included a coarse correction for chemical peculiarities, based mostly on magnetic Ap stars,
they did not tailor their models to the specifics of the (HgMn) AR Aur system and used only the photometric temperature indicators. 

We adopt the \lgg(A)\,=\,4.33 and \lgg(B)\,=\,4.28 of \citet{Nordstrom1994-arAur-fundamental}, based on their precise masses and radii. As noted by these authors, the observation that \lgg(A)\,$>$\,\lgg(B) and $R_A < R_B$ suggests that the primary is a ZAMS star while the secondary is still contracting toward the ZAMS.  However, this conclusion relies on the assumption of identical Mg abundances in the two stars, which may not be fulfilled since both components of AR Aur are chemically peculiar.  
The Hertzsprung-Russell diagram positions of the two stars are consistent with this conclusion, 
however they do not rule out both stars being at the ZAMS.

\begin{figure}
\centering
\includegraphics[width=3.2in]{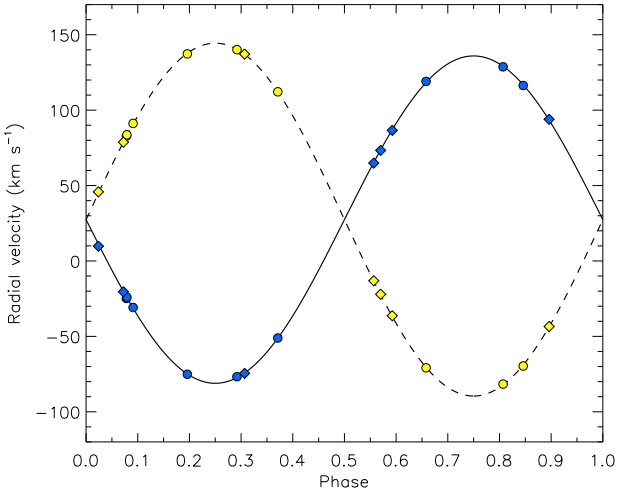}
\caption{The orbital variation of radial velocities for AR~Aur~A and B. Symbols show measurements obtained with UVES (circles) and ESPaDOnS/NARVAL (diamonds) for the primary (dark symbols) and secondary (light symbols). The lines show fitted radial velocity curves for the primary (solid line) and secondary (dashed line).}
\label{orbit}
\end{figure}

The lack of an accurate spectroscopic orbital solution based on precise radial velocity measurements of AR~Aur has prevented inclusion of this system in the list of fundamental eclipsing binary stars \citep*{Torres2010}. One can expect that new high-quality spectroscopy of the AR~Aur system presented by \citet{Hubrig2006-arAur-lVar} and in our paper might improve on this situation. A combined set of radial velocities, determined from our spectra and taken from Fig.~2 of Hubrig et al., is given in Table~\ref{rvs}. The typical uncertainty of these measurements is 0.5~\kms. 
We modelled the radial velocity curves of both components adopting the period from \citet{Albayrak2003} 
and assuming a circular orbit. We found it necessary to correct our radial velocity measurements by 1.32~\kms\ 
to match the zero-point of Hubrig et al.  A small phase shift of $0.0024\pm0.0003$ with respect to the ephemeris 
of Albayrak et al. is also required to account for the light-time effect caused by the third component. 
The magnitude and sign of this shift are consistent with the predictions by \citet{Albayrak2003} and \citet{Chochol2006}. 
The resulting fits are presented in Fig.~\ref{orbit}. Our semi-amplitudes, $K_A=108.54\pm0.21$~\kms\ and $K_B=116.99\pm0.21$~\kms, yield a mass ratio of $M_A/M_B=1.078\pm0.003$ and masses of the individual components $M_A=2.552\pm0.008$~\Msun\ and $M_B=2.367\pm0.008$~\Msun\ for the orbital inclination of $i=88.52\degr\pm0.06\degr$ \citep{Nordstrom1994-arAur-fundamental}.

\begin{table}
\centering
\caption{Radial velocity measurements of AR Aur.  Typical uncertainties in the radial velocity are 0.3-0.5 \kms. 
Phases were computed using the ephemeris of \citet{Albayrak2003}, given in Sect.~\ref{observations}.  }
\label{rvs}
\begin{tabular}{cccrr}
\hline \hline \noalign{\smallskip}
HJD & Instrument & Phase & $V_A$~~~ & $V_B$~~~ \\
$-2400000$ & & & (\kms) & (\kms) \\
\noalign{\smallskip} \hline \noalign{\smallskip}
 53671.825 &     UVES &  0.078 & $-24.86$ & $ 82.74$ \\
 53700.773 &     UVES &  0.079 & $-24.06$ & $ 83.66$ \\
 53725.627 &     UVES &  0.091 & $-30.86$ & $ 91.20$ \\
 53713.659 &     UVES &  0.196 & $-75.14$ & $137.31$ \\
 53730.594 &     UVES &  0.292 & $-76.74$ & $140.11$ \\
 53722.655 &     UVES &  0.371 & $-51.09$ & $112.17$ \\
 53632.872 &     UVES &  0.658 & $119.09$ & $-70.86$ \\
 53728.592 &     UVES &  0.807 & $128.74$ & $-81.66$ \\
 53724.619 &     UVES &  0.846 & $116.34$ & $-69.66$ \\
 54070.874 & ESPaDOnS &  0.590 & $ 87.93$ & $-35.01$ \\
 54074.915 & ESPaDOnS &  0.568 & $ 74.64$ & $-20.73$ \\
 54076.792 & ESPaDOnS &  0.021 & $ 11.12$ & $ 47.13$ \\
 54084.533 &   NARVAL &  0.894 & $ 95.25$ & $-42.15$ \\
 54089.396 &   NARVAL &  0.070 & $-19.09$ & $ 80.09$ \\
 54090.366 &   NARVAL &  0.304 & $-73.21$ & $138.50$ \\
 54091.399 &   NARVAL &  0.554 & $ 66.22$ & $-11.81$ \\
\noalign{\smallskip} \hline \hline
\end{tabular}
\end{table}

\section{Chemical Abundances}
\label{Chemical Abundances}
A detailed abundance analysis was preformed for both AR~Aur~A and B, using the disentangled spectra.  
The ZEEMAN2 spectrum synthesis code \citep{Landstreet1988-Zeeman1,Wade2001-zeeman2_etc} was used, 
which solves the polarized radiative transfer equations, assuming LTE. 
Optimizations to the code for negligible magnetic fields values were included, 
and a Levenberg-Marquardt $\chi^2$ minimization routine was used to aid in fitting the observed spectrum.  

Atomic data was extracted from the Vienna Atomic Line Database (VALD) \citep{Kupka1999-VALD}, 
using an {\sc extract stellar} request, with chemical abundances tailored to the two stars, 
based on the results of \citet{Ryabchikova1998-HgMn-binaries}.  
The model atmospheres calculated for the fundamental parameters derived
in Sect.~\ref{params} using the LLModels code were adopted in the abundance analysis.

Prior to abundance analysis, the average disentangled spectra of the primary and secondary were corrected for 
the continuum dilution by the other component \citep[see][]{Folsom2008} using a theoretical, wavelength dependent light ratio 
predicted by the adopted LLModels atmospheres.

The spectra of both stars were fit for chemical abundances, as well as \vs\, and microturbulence.  
Fitting was performed for 11 independent regions, between 100 and 200~\AA\, long, 
ranging from 4170~\AA\, to 6200~\AA.  
The chemical abundances used as free parameters in each spectral window were chosen by
comparing the observation to the atomic data by eye, 
and looking for lines with corresponding features in the observed spectrum.  
This was then checked using synthetic spectra to ensure a good constraint was available. 
Final best fit values are averages of the results 
over the 11 windows, with the window-to-window standard deviation of the abundance adopted as the experimental uncertainty. 
The final best fit averages and uncertainties are presented in Table~\ref{abundance-tab}. 
In cases where less than four lines are available the uncertainty estimate was made by eye, 
including potential normalization errors, blended lines, and the scatter between lines.  
The number of lines of atomic data that contributed significantly to the best fit abundances are 
given in Table~\ref{abundance-tab} in brackets.  
Additional very weak lines were included to provide an accurate spectral synthesis, 
but such lines contribute very little to the fit and the derived abundances, and 
have been excluded from these numbers.

\begin{figure*}
\centering
\includegraphics[width=6.5in]{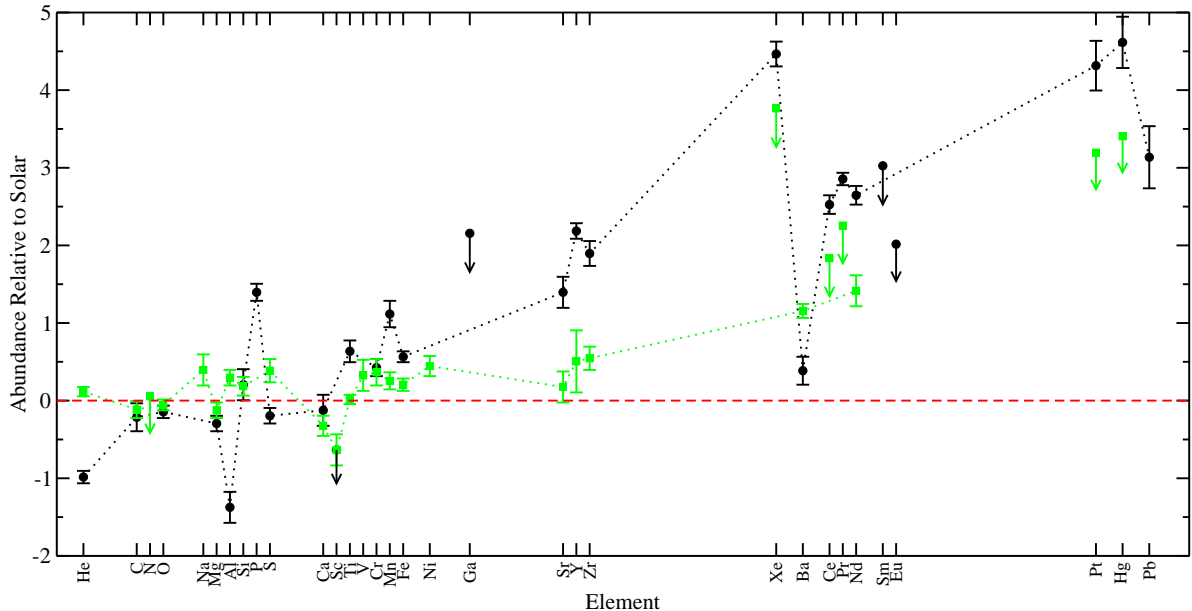}
\caption{Abundances relative to solar for AR~Aur~A (circles) 
and AR~Aur~B (squares), averaged over all spectral windows modeled.  
Solar abundances are taken from \citet{2005Asplund-solar-abun}.  
Points marked with an arrow represent upper limits only.  }
\label{abunplot}
\end{figure*}

\begin{table}
\centering
\caption{Averaged best fit chemical abundances, \vs\, and microturbulence ($\xi$) for AR~Aur~A and B as 
well as solar abundances from \citet*{2005Asplund-solar-abun}. 
Values given in brackets are the number of lines of atomic data on which the abundances are based. 
Abundances are given in units of $\log\left(\frac{N_{\rm X}}{N_{\rm tot}}\right)$.}
\begin{tabular}{cc@{ }cc@{ }cc}
\hline \hline \noalign{\smallskip}
        &\multicolumn{2}{c}{AR Aur A}&\multicolumn{2}{c}{AR Aur B} & Solar \\
\noalign{\smallskip} \hline \noalign{\smallskip}
\vs &\multicolumn{2}{c}{$23.1\pm0.9$ \kms}&\multicolumn{2}{c}{$22.9\pm0.7$ \kms}&  \\
$\xi$ &\multicolumn{2}{c}{$\leq 1$ \kms}    &\multicolumn{2}{c}{$\leq 1$ \kms}    &  \\
\noalign{\smallskip} \hline \noalign{\smallskip}
He      & $-2.09\pm0.08$  & $(3  )$ & $-0.99\pm0.06$ & $(4  )$ & $ -1.11$ \\
C       & $-3.86\pm0.18$  & $(5  )$ & $-3.76\pm0.11$ & $(5  )$ & $ -3.65$ \\
N       & 	          &         & $\leq -4.2$    & $(1  )$ & $ -4.26$ \\
O       & $-3.52\pm0.08$  & $(4  )$ & $-3.43\pm0.07$ & $(4  )$ & $ -3.38$ \\
Na      & 	          &         & $-5.47\pm0.20$ & $(2  )$ & $ -5.87$ \\
Mg      & $-4.80\pm0.10$  & $(8  )$ & $-4.63\pm0.10$ & $(12 )$ & $ -4.51$ \\
Al      & $-7.04\pm0.20$  & $(1  )$ & $-5.37\pm0.10$ & $(2  )$ & $ -5.67$ \\
Si      & $-4.32\pm0.20$  & $(8  )$ & $-4.34\pm0.12$ & $(8  )$ & $ -4.53$ \\
P       & $-5.28\pm0.11$  & $(4  )$ &                &         & $ -6.68$ \\
S       & $-5.09\pm0.10$  & $(2  )$ & $-4.51\pm0.15$ & $(4  )$ & $ -4.90$ \\
Ca      & $-5.85\pm0.20$  & $(3  )$ & $-6.05\pm0.13$ & $(6  )$ & $ -5.73$ \\
Sc      & $\leq -9.5$     & $(1  )$ & $-9.50\pm0.20$ & $(6  )$ & $ -8.87$ \\
Ti      & $-6.50\pm0.14$  & $(45 )$ & $-7.12\pm0.06$ & $(31 )$ & $ -7.14$ \\
V       & 	          &         & $-7.71\pm0.20$ & $(4  )$ & $ -8.04$ \\
Cr      & $-5.97\pm0.11$  & $(32 )$ & $-6.03\pm0.17$ & $(43 )$ & $ -6.40$ \\
Mn      & $-5.53\pm0.17$  & $(43 )$ & $-6.39\pm0.11$ & $(12 )$ & $ -6.65$ \\
Fe      & $-4.02\pm0.07$  & $(356)$ & $-4.38\pm0.08$ & $(200)$ & $ -4.59$ \\
Ni      & 	 	  &         & $-5.36\pm0.13$ & $(10 )$ & $ -5.81$ \\
Ga      & $\leq -7.0$	  & $(1  )$ &                &         & $ -9.16$ \\
Sr      & $-7.72\pm0.20$  & $(1  )$ & $-8.94\pm0.20$ & $(1  )$ & $ -9.12$ \\
Y       & $-7.64\pm0.10$  & $(23 )$ & $-9.32\pm0.40$ & $(3  )$ & $ -9.83$ \\
Zr      & $-7.58\pm0.16$  & $(26 )$ & $-8.93\pm0.15$ & $(1  )$ & $ -9.48$ \\
Xe      & $-5.30\pm0.16$  & $(4  )$ & $\leq -6.0$    & $(2  )$ & $ -9.77$ \\
Ba      & $-9.48\pm0.18$  & $(1  )$ & $-8.71\pm0.09$ & $(3  )$ & $ -9.87$ \\
Ce      & $-7.81\pm0.12$  & $(27 )$ & $\leq -8.5$    & $(3  )$ & $-10.34$ \\
Pr      & $-8.60\pm0.08$  & $(6  )$ & $\leq -9.2$    & $(3  )$ & $-11.46$ \\
Nd      & $-7.94\pm0.12$  & $(21 )$ & $-9.18\pm0.20$ & $(2  )$ & $-10.59$ \\
Sm      & $\leq -8.0$     & $(3  )$ &                &         & $-11.03$ \\
Eu      & $\leq -9.5$     & $(2  )$ &                &         & $-11.52$ \\
Pt      & $-6.08\pm0.32$  & $(3  )$ & $\leq -7.2$    & $(1  )$ & $-10.40$ \\
Hg      & $-6.29\pm0.33$  & $(4  )$ & $\leq -7.5$    & $(1  )$ & $-10.91$ \\
Pb      & $-6.9\pm0.4$    & $(2  )$ &                &         & $-10.04$ \\

\noalign{\smallskip} \hline \hline
\end{tabular}
\label{abundance-tab}
\end{table}

\subsection{Abundances in AR Aur A}
\begin{figure*}
\centering
\includegraphics[width=5.0in]{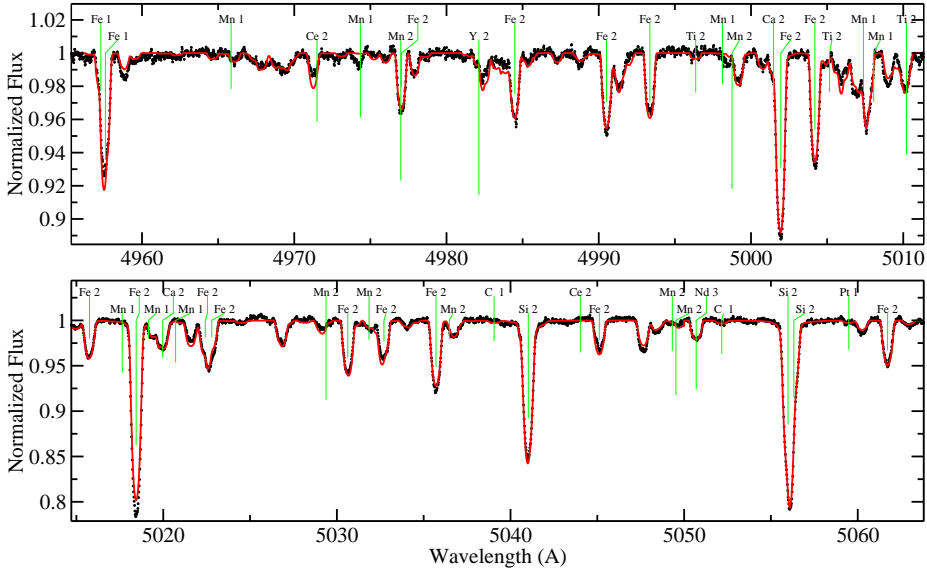}
\caption{Sample fits to the spectrum of AR~Aur~A.  
Points represent the observation and the smooth line represents the best fit spectrum.  
Elements which are major contributors to each line have been labeled.  }
\label{fit-primary}
\end{figure*}

The best fit synthetic spectrum for AR~Aur~A generally reproduces the observed spectrum quite well.  
Sample fits to the spectrum of AR~Aur~A are shown in Fig.~\ref{fit-primary}.  
The average best fit parameters are shown in Table~\ref{abundance-tab}, 
and chemical abundances are plotted relative to solar values in Fig.~\ref{abunplot}.  
The effect of hyper-fine splitting was examined for Mn lines and generally found to be small. 
In the few cases where hyperfine splitting became important, the line was excluded from our fit.

AR~Aur~A is clearly an HgMn star, with dramatic overabundances of Hg and Mn.  
The Fe-peak elements are significantly enhanced, as are P, Sr, Y and Zr.  
Several rare earths, Xe, Pt and Pb are also strongly enhanced.  
C and O are approximately solar and Ba appears to be near solar as well.  
He is significantly underabundant, as is Al.  
Only an upper limit on the microturbulence of 1~\kms\, could be derived for this star, 
and abundances here have been calculated assuming no microturbulence.  
A microturbulence of 1~\kms\, would decrease the abundances derived by approximately 0.05 dex. 
This upper limit on microturbulence is based on the window-to-window scatter of best fit values, 
when microturbulence was included as a free parameter in the $\chi^2$ fit. 
The majority of spectral windows produced a best fit value of 0~\kms, 
though some windows produced marginally non-zero values. 
Thus the `best' overall value of 0~\kms\, was adopted, and the fitting procedure was repeated using this value. 

The abundance we derive generally agree with those determined by \citet{Ryabchikova1998-HgMn-binaries}.  
The exceptions to this are C, Mn, Sr, Y, and Pt.  For these five elements we find lower abundances than 
\citet{Ryabchikova1998-HgMn-binaries} by between 0.5 and 1 dex, 
which is outside of our uncertainty and any likely uncertainty in her results.  
For Sr, Y, and Pt strong line profile variability, discussed below (Sect.~\ref{lpv}), could be responsible for
some of this discrepancy.

\subsection{Abundances in AR Aur B}
\begin{figure*}
\centering
\includegraphics[width=5.0in]{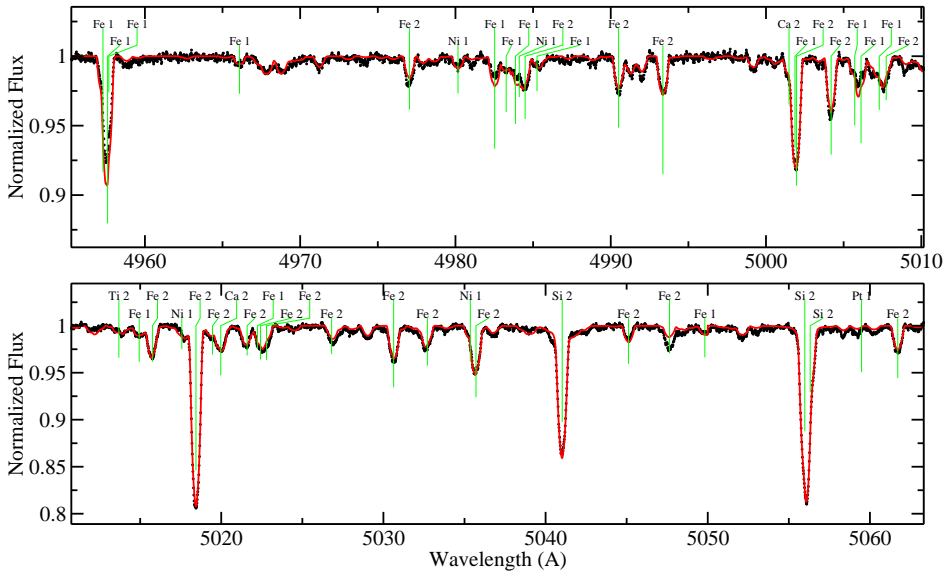}
\caption{Sample fits to the spectrum of AR~Aur~B.  
Points represent the observation and the smooth line represents the best fit spectrum.  
Elements which are major contributors to each line have been labeled.  }
\label{fit-secondary}
\end{figure*}

For AR Aur B, we again achieve a very good fit to the observations, 
as presented in Fig.~\ref{fit-secondary}.  The best fit abundances are shown 
in Table~\ref{abundance-tab}, and plotted relative to solar in Fig.~\ref{abunplot}.

Based on these abundances, AR~Aur~B appears to be a very weak Am star.  
The Fe peak elements are, on average, enhanced by 0.3 dex, while He, C, and O are all nearly solar.  
Ca and Sc are both significantly underabundant, while Ba and Nd are both substantially overabundant.  
Again, an upper limit on the microturbulence of 1~\kms\, was derived, 
and the abundances are based on a 0~\kms\, model.  
This limit was determined in the same fashion as for AR~Aur~A.  
The underabundant Ca coupled with the overabundant Fe-peak elements, Ba, and Nd 
strongly suggest that this is a weak Am star.  

We find a good agreement between our abundances and those derived by \citet{Ryabchikova1998-HgMn-binaries} 
for the majority of elements.  However we find lower abundances for Ti, Mn, Fe, and possibly Ba, 
beyond our uncertainties and the likely uncertainties of \citet{Ryabchikova1998-HgMn-binaries}.  
Due to the higher S/N and comparatively advanced spectral disentangling used in our observations, 
we consider our results to be more accurate.  

\section{Line Profile Variability}
\label{lpv}
We see clear variability in a number of lines in our disentangled spectrum of AR Aur A.  
Line variability was diagnosed by examining regions with an abnormally large 
deviation between the observations and the model spectra produced by the disentangling procedure.  
Regions in which a large deviation was seen throughout most of a line profile were considered significant. 
Regions of deviation outside of a line, or in lines with only a small portion deviating significantly, 
were considered to be spurious.  
In the spectrum of AR Aur A such spurious regions are rare, and are almost always attributable to the edges of 
spectral orders or cosmic rays/bad pixels.  Regions with heavy telluric line contamination were avoided entirely. 
This method is superior to that employed in previous studies in that it provides a quantitative measure of variability, 
in the form of the standard deviation spectra, rather then a simple estimate by eye.  

\begin{figure*}
\centering
\includegraphics[width=2.2in,angle=90]{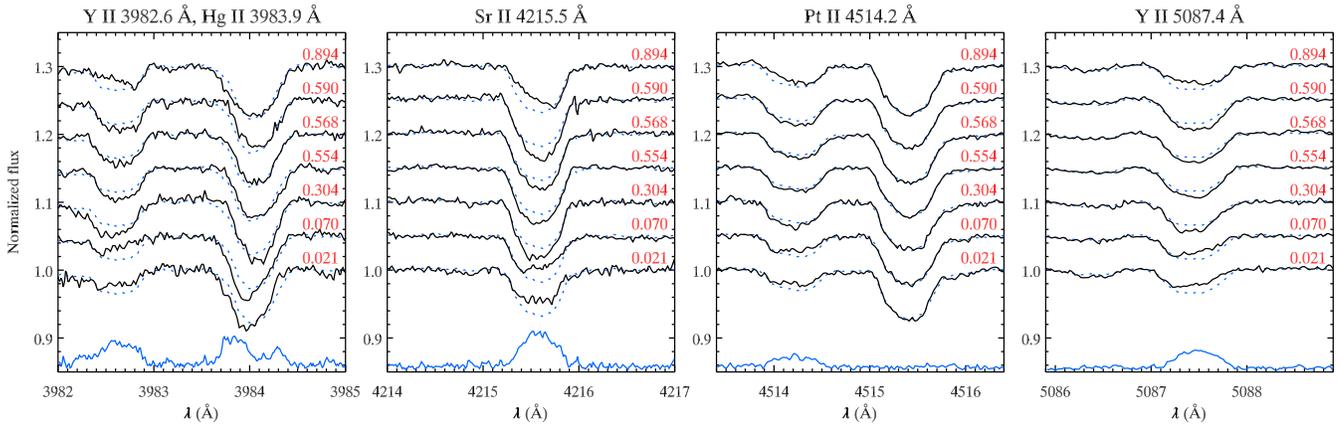}
\caption{Line profile variability in the spectrum of AR~Aur~A. The spectra are plotted for Y~{\sc ii} 3982.5~\AA,
Hg~{\sc ii} 3983.9~\AA, Sr~{\sc ii} 4215.5~\AA, Pt~{\sc ii} 4514.2~\AA, and Y~{\sc ii} 5087.4~\AA. Disentangled
observations for different phases (solid lines) are shifted vertically. The dotted line gives the average spectrum.
The standard deviation profile, shifted upwards and scaled up by a factor of 3, is shown at the bottom of each panel.}
\label{lpv1}
\end{figure*}

Clear variability was seen in lines of Cr, Mn, Sr, Y, Ba, Pt, and Hg in the primary (see Fig.~\ref{lpv1}).
These are some of the more strongly overabundant elements in the star, with the exception of Ba.  
Lines with observed variability in the primary are listed in Table~\ref{var-lines}.  
The lines in which we observe variability are generally the stronger lines 
of the elements in question.  This suggests that all lines of these elements are in fact variable, 
and the cases in which we observe no variability simply fall below our detection threshold.  

The rotation and orbital motion of the AR~Aur components is believed to be synchronized 
\citep{Nordstrom1994-arAur-fundamental,Khokhlova1995-araur-ew}. Thus, we presume that 
the line profile variations we observe occur with the period of 4.135~d. Although we cannot derive the line variability 
period independently, a good agreement between profiles obtained at the same orbital phase two
weeks apart confirm the assumption of synchronous rotation.
Since particular variability pattern is restricted to specific elements (e.g., the variability of Hg and Y lines is clear different), 
we conclude that these spectral changes are produced by  
an inhomogeneous surface distribution of the elements in question.

\begin{figure}
\centering
\includegraphics[width=2.0in,angle=90]{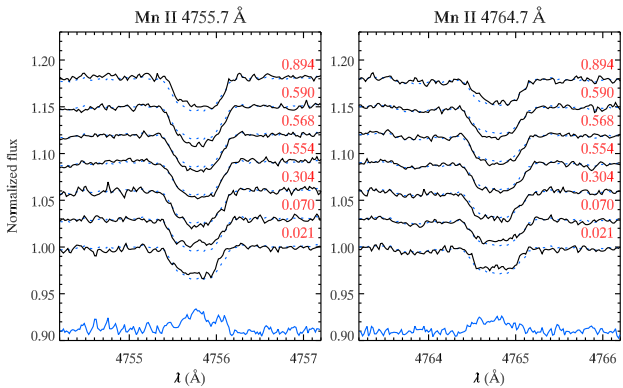}
\caption{Weak profile variability in the Mn~{\sc ii} 4755.7 and 4764.7~\AA\ lines. 
The format of this figure is similar to Fig.~\ref{lpv1}. The standard deviation profile is scaled up by a factor of 5.}
\label{lpv2}
\end{figure}

In comparison with \citet{Hubrig2006-arAur-lVar}, we confirm their detection of variability in 
Y, Pt, Hg, and Sr, but we see no evidence for variability in O, Na, Mg, Si, C, Ti, Fe, He, Nd, and Zr.  
Additionally, we see evidence for variability in Cr, Mn, and Ba which these authors did not note. 
An example of weak variations seen in the Mn~{\sc ii} lines is presented in Fig.~\ref{lpv2}.
A full comparison of specific lines in which variability is observed was not possible, 
since \citet{Hubrig2006-arAur-lVar} did not publish such a list.  However, in the lines they did mention, 
we see variability in the Pt~{\sc ii} 4061.7, Hg~{\sc ii} 3983.9, Sr~{\sc ii} 4077.7, 
and Y~{\sc ii} 4900.1 lines, 
but we see no clear variability in the He~{\sc i} 5015.7 and Nd~{\sc iii} 4927.5 lines.
In the cases where \citet{Hubrig2006-arAur-lVar} report variability that we do not observe, 
it may be that the variability simply falls below our detection threshold, 
though this would imply a very low amplitude of variability, or that we observed 
the system at unfavourable phases for detecting variations.  
However it is also possible that some of the variability reported 
is due to spectral features of the secondary that have not been completely removed, 
or contamination by weak telluric lines.  

\begin{table}
\centering
\caption{Variable lines observed in AR Aur A.  If a line is blended significantly 
multiple components are listed.  Lines marked by a question mark show only 
very weak variability, near the noise level. }
\begin{tabular}{cccccc}
\hline \hline \noalign{\smallskip}
 Ion        & Wavelength & & Ion      & Wavelength & Weak \\
            & (\AA)      & &            & (\AA)    &      \\
\noalign{\smallskip} \hline \noalign{\smallskip}
 Y  {\sc ii} &   3950.352 & &            &          &   \\
 Y  {\sc ii} &   3982.594 & &		&	   &   \\
 Hg {\sc ii} &   3983.931 & &		&	   &   \\
 Pt {\sc ii} &   4046.449 &+& Hg {\sc i}  & 4046.559 &   \\
 Pt {\sc ii} &   4061.659 & &		&	   &   \\
 Sr {\sc ii} &   4077.709 & &		&          &   \\
 Y  {\sc ii} &   4124.907 & &            &          & ? \\
 Sr {\sc ii} &   4161.792 & &	        &	   &   \\
 Y  {\sc ii} &   4177.529 & &	        &	   &   \\
 Y  {\sc ii} &   4204.695 & &            &          & ? \\
 Mn {\sc ii} &   4206.367 & &            &          & ? \\
 Sr {\sc ii} &   4215.519 & &		&	   &   \\
 Y  {\sc ii} &   4235.729 & &		&	   &   \\
 Mn {\sc ii} &   4242.333 &+& Cr {\sc ii} & 4242.364 & ? \\
 Sr {\sc ii} &   4305.443 & &		&	   &   \\
 Y  {\sc ii} &   4309.631 & &		&	   &   \\
 Mn {\sc ii} &   4326.639 & &            &          & ? \\
 Hg {\sc i}  &   4358.323 &+& Y  {\sc ii} & 4358.728 & ? \\
 Y  {\sc ii} &   4374.935 & &		&	   &   \\
 Y  {\sc ii} &   4398.013 & &		&	   &   \\
 Y  {\sc ii} &   4422.591 & &		&	   &   \\
 Mn {\sc ii} &   4478.637 & &            &          & ? \\
 Pt {\sc ii} &   4514.170 & &	       	&          &   \\
 Ba {\sc ii} &   4554.029 & &            &          & ? \\
 Cr {\sc ii} &   4554.988 & &            &          & ? \\
 Cr {\sc ii} &   4558.650 & &            &          & ? \\
 Cr {\sc ii} &   4618.803 & &            &          & ? \\
 Cr {\sc ii} &   4634.070 & &            &          & ? \\
 Y  {\sc ii} &   4682.324 & &            &          &   \\
 Mn {\sc ii} &   4755.727 & &            &          & ? \\
 Mn {\sc ii} &   4764.728 & &            &          & ? \\
 Y  {\sc ii} &   4786.580 & &            &          & ? \\
 Y  {\sc ii} &   4823.304 &+& Mn {\sc i}  & 4823.524 &   \\
 Cr {\sc ii} &   4876.399 & &            &          & ? \\
 Y  {\sc ii} &   4883.684 & &		&	   &   \\
 Y  {\sc ii} &   4900.120 & &		&	   &   \\
 Y  {\sc ii} &   5087.416 & &		&	   &   \\
 Y  {\sc ii} &   5123.211 &+& Mn {\sc ii} & 5123.327 &   \\
 Y  {\sc ii} &   5200.406 & &		&          &   \\
 Y  {\sc ii} &   5205.724 & &		&	   &   \\
 Cr {\sc ii} &   5237.329 & &            &          & ? \\
 Mn {\sc ii} &   5297.028 &+& Mn {\sc ii} & 5297.056 & ? \\
 Mn {\sc ii} &   5299.330 &+& Mn {\sc ii} & 5299.386 & ? \\
 Y  {\sc ii} &   5402.774 & &		&	   &   \\
 Y  {\sc ii} &   5497.408 & &		&	   &   \\
 Y  {\sc ii} &   5509.895 & &		&	   &   \\
 Y  {\sc ii} &   5662.925 & &            &          &   \\ 
\noalign{\smallskip} \hline \hline
\end{tabular}
\label{var-lines}
\end{table}

\section{Magnetic Fields}

The presence (or absence) of magnetic fields in the photospheres of AR Aur A/B is diagnosed from 
the presence of circular polarisation within spectral lines in the ESPaDOnS and NARVAL spectra, produced as a consequence of the longitudinal Zeeman effect. 

No circular polarisation signatures were observed in any of the individual spectral lines of the primary or secondary. We therefore used Least Squares Deconvolution (LSD) \citep{Donati1997-major} to extract mean Stokes $I$, $V$ and diagnostic $N$ profiles from the 7 individual observed spectra, and to thereby increase the sensitivity of the magnetic diagnosis. 

\begin{figure}
\centering
\includegraphics[width=7.3cm,angle=-90]{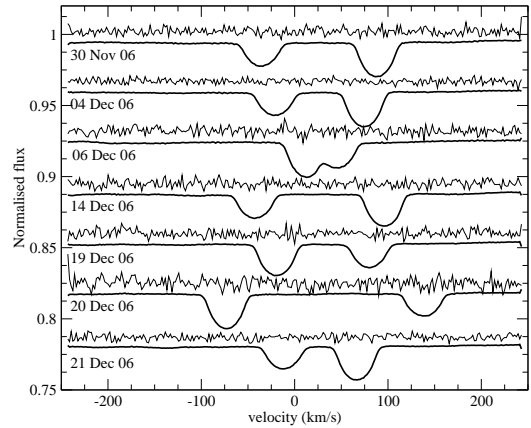}
\caption{LSD Stokes $I$ (lower) and $V$ (upper) profiles extracted from the 7 observations of the AR Aur system. 
The primary and secondary components are distinguished by their slightly different line depths. 
The Stokes $V$ profile amplitudes are scaled up by 25 times with respect to $I$ for clarity.  }
\label{lsdprofs}
\end{figure}

The LSD analysis was performed using the original reduced spectra representing the combined system. 
Spectral disentangling was not possible for the Stokes $V$ spectra since the disentangling technique relies on both detectable lines, which we do not have in Stokes $V$, 
and lines remaining roughly constant, which would not be the case in Stokes $V$ if they were detected.  
The line mask used for the LSD process was constructed using {\sc extract stellar} requests from VALD, %the Vienna Atomic Line Database (VALD), 
for the atmospheric parameters and abundances {of AR Aur A}\ derived in Sect.~\ref{params} and \ref{Chemical Abundances}. The mask was filtered to include only lines with predicted unbroadened depths great than 10\% of the continuum, and ultimately contained 1168 lines.  As the atmospheric parameters and Fe-peak element abundances of AR Aur A and B are similar, and because such lines represent the majority of the lines used in the mask, the use of the primary abundances for the LSD extraction was not expected to be a significant source of error. This was confirmed by experiment. The Stokes $I$ and $V$ LSD profiles are illustrated in Fig.~\ref{lsdprofs}

\begin{table*}
\centering
\caption{Longitudinal magnetic field measurements of AR Aur. Columns report the UT date of observation, signal-to-noise ratio per 1.8 km/s LSD pixel in the extracted LSD $N$ profile. Then, for both the primary and secondary respectively: Stokes $V$ detection diagnosis (ND=no detection), longitudinal field measured using $V$ and $N$ across the mean line in units of Gauss. On 6 Dec 2006, the profiles of AR Aur A and B were blended together. The longitudinal field measured from the combined profiles was $57\pm 33$~G (Stokes $V$) and $18\pm 33$~G (diagnostic null).}
\label{magnetic_table}
\begin{tabular}{cclrr|crr}
\hline \hline \noalign{\smallskip}
          & & \multicolumn{3}{c}{AR Aur A} & \multicolumn{3}{c}{AR Aur B}\\
Date & LSD S/N& Det & $V\ B_\ell$ (G) & $N\ B_\ell$ (G) & Det & $V\ B_\ell$ (G) & $N\  B_\ell$ (G)\\
\noalign{\smallskip} \hline \noalign{\smallskip}
30 Nov 06& 10800       & ND    &    $ -29\pm      26   $         &  $-24   \pm     26$    & ND &     $ 50\pm      43 $   &     $-77\pm      43$  \\
04 Dec 06    &  14200       & ND   &  $ 29\pm      19     $     &   $-31   \pm     19  $    & ND   &   $  -24\pm      29  $   &  $-10\pm      29$ \\  
14 Dec 06        &   10500     & ND  &  $ -10\pm      31   $        &   $-3   \pm     31   $   & ND      &  $-31\pm      36  $    &  $-19\pm      36$\\  
19 Dec 06    &   10500      & ND  &  $  2\pm      34    $        &    $-44  \pm      34  $  & ND    &  $ 33\pm      48    $   &  $ 27\pm      49$ \\  
20 Dec 06      &     7100    & ND  & $  9\pm      42     $        &    $32   \pm     43  $   & ND   &  $ 12\pm      69    $    & $ 15\pm      70 $ \\  
21 Dec 06        &  13400     & ND    & $ -14\pm      23    $        &  $-14   \pm     23 $    & ND   & $ -26\pm      36   $   &  $ -23\pm      36 $\\ 
\noalign{\smallskip} \hline \hline
\end{tabular}
\end{table*}

The LSD profiles are characterised by signal-to-noise ratios (S/N; per 1.8 km/s pixel) ranging from about 7000 to about 14000, corresponding to a net S/N gain of about 20 times. No signal was detected in the $V$ or $N$ LSD spectra of any of the observations, 
based on the detection criteria of  \citet{Donati1997-major}.  The longitudinal magnetic field was derived from each observation using Eq. (1) of \citet{Wade2000-highPrecision-correctBz}. 
To derive the longitudinal field of each of the components of AR Aur, we integrated across the line profiles of each star at phases where those profiles were separated in velocity and unblended. The integration range was carefully established by eye for each profile; the full integration range was typically $30\pm 1$~km/s on either side of the profile center-of-gravity. No statistically-significant (i.e. greater than 3$\sigma$) longitudinal field was detected in either star, with typical $1\sigma$ uncertainties of 30--40 G. The results of the magnetic analysis are reported in Table~\ref{magnetic_table}. We point out that the inferred value of the longitudinal field of a component of a binary system is unaffected by the continuum of the companion. This is because it is the continuum normalised Stokes $I$ and $V$ parameters ($I/I_{\rm c}$ and $V/I_{\rm c}$, respectively) that appear in the denominator and numerator of Wade et al.'s Eq. (1). 
Nevertheless, the inferred uncertainty, derived from photon-noise error bars propagated through the LSD procedure, is larger than that for the single-star case.
  
From these basic magnetic measurements, we conclude that neither component hosts a magnetic field with a disc-averaged longitudinal component larger than about 100~G. If either star hosted a field with a predominantly dipolar topology (like those of the magnetic Ap stars), the individual measurements indicate that the polar surface strength of that field is roughly constrained to be weaker than about 300~G.

Because the stellar components of the AR Aur system are likely to be tidally locked with their rotation, we can phase the longitudinal field measurements and LSD profiles according to the orbital (and therefore presumably rotational) ephemeris give in Sect.~2. We point out that the orbital period of 4.1346656 days, interpreted as a rotational period, is in excellent agreement with the radii and $v\sin i$s of the components, assuming that the inclination of the stellar rotational axes is equal to the orbital inclination.

We have phased the longitudinal field measurements, and examined their agreement with synthetic longitudinal field curves corresponding to a large grid of dipole surface magnetic field models. 
While three of the measurements occur at approximately the same phase, the dataset samples the entire rotational cycle at approximately quarter-cycle intervals. This sparse but relatively uniform sampling is important for constraining the field intensity using longitudinal field measurements. 
We assume the stellar rotation axis inclination $i\simeq 88.5\degr$, in agreement with the orbital inclination and consistent with our assumption that the system is tidally locked. We find for perpendicular ($\beta\simeq 90\degr$) dipoles, the 6 measurements constrain the polar strength of any dipole magnetic field in the primary to below 225~G, at $3\sigma$ confidence. For intermediate field obliquities ($\beta\simeq 45\degr$), the upper limit on any dipole field present is about 375~G. For the secondary, the analogous upper limits are similar: about 275~G and 400~G, respectively. 

The lack of detection of any signature in the Stokes $V$ profiles also constrains the presence of more complex magnetic fields in the photospheres of AR Aur A/B. However, to place quantitative upper limits on the presence of such fields requires an {\em ad hoc} assumption about their structure and intensity. Such an investigation is beyond the scope of this paper. 

In summary, our observations provide no evidence for the presence of magnetic fields in either of the components of AR Aur. 
If oblique dipolar fields are present in either star, the polar strength of those fields are constrained to be weaker than a few hundred G at 3$\sigma$ confidence. 
As fields of intensity similar to our upper limit \citep[e.g. 300--400~G in $\epsilon$~UMa;][]{Bohlender1990-epsUMa} 
are capable of producing strong chemical nonuniformities in the atmospheres of Ap stars, we are not able to fully rule out a magnetic origin of these features. 
Nevertheless, such weak fields in Ap stars appear to be quite rare - \citet{Auriere2007} 
identify only 3 stars out of a modeled sample of 24 stars with fields that are inferred to be similarly weak. 
Therefore, given the rarity of such fields, and the high ($3\sigma$) confidence of our null result, 
we conclude that it is highly probable that the chemical nonuniformities inferred to exist in the atmosphere of AR Aur A are not of magnetic origin, 
and are produced by some other phenomenon.

\section{Conclusions}

We find AR Aur A to be a strongly peculiar HgMn star, 
confirming previous results \citep{Khokhlova1995-araur-ew,Ryabchikova1998-HgMn-binaries}.  
AR Aur B shows weak Am peculiarities, particularly modest overabundances of Fe-peak elements and Nd, 
and modest underabundances of Ca and Sc.  Thus AR Aur is a binary system with two different chemically peculiar stars.  
This is striking in light of the very similar temperatures of the two stars.  
It is possible that the small temperature difference represents a sharp dividing line between Am and HgMn stars.  
Alternately, if AR Aur B is still on the pre-main sequence, 
the difference in chemical abundances may be a result of the different evolutionary status of the stars. 

We confirm the result of \citet{Nordstrom1994-arAur-fundamental} that the cooler, 
lower mass, secondary has a larger radius than the primary, 
assuming that both components have the same Mg abundance.  
Under this assumption, we agree with their conclusion that 
this implies the secondary is still contracting towards the ZAMS, 
while the primary has likely already arrived on the ZAMS.  Thus both components of AR Aur are 
likely some of the youngest chemically peculiar stars known.  
AR Aur B may be the first pre-main sequence Am star discovered.  
This result, if correct, implies that chemical peculiarities must arise quickly 
in both HgMn stars and in Am stars.  The atomic diffusion process thought to drive 
the observed chemical peculiarities must be fairly efficient for this to occur.  

Despite a careful high S/N search, we find no evidence for a magnetic field in AR Aur A or B.  
The LSD profiles for our observations, which are sensitive to simpler field geometries 
with a zero net longitudinal component, as well as more complex field geometries, 
show no indication of any signal in either star.  
Our longitudinal field measurements place a $3\sigma$ upper limit of 100~G on the longitudinal fields of both 
stars over a range of rotational phases. 
This constrains a dipole field to be less than about 400 G at the pole in either star.  
Thus we conclude that neither AR Aur A nor B is likely to have any significant magnetic field.  
This matches the result of \citet{Wade2006-alphaAnd} for the HgMn star $\alpha$~And.

AR Aur represents a second case in which line profile variability and 
surface abundance inhomogeneities exist in the absence of any strong magnetic field.  
This substantially strengthens the conclusion that, 
whatever the mechanism producing these inhomogeneities, a magnetic field is not required.

\section*{Acknowledgments} 
OK is a Royal Swedish Academy of Sciences Research Fellow supported by grants from the Knut and Alice Wallenberg Foundation and the Swedish Research Council.
GAW is supported by an Natural Science and Engineering Research Council (NSERC Canada) Discovery Grant and a Department of National Defence (Canada) ARP grant.

\bibliography{masivebib.bib}{}
\bibliographystyle{mn2e}

\label{lastpage}

\end{document}